\documentclass[conference]{IEEEtran}

\usepackage{cite}
\usepackage{amsmath,amssymb,amsfonts}
\usepackage{algorithmic}
\usepackage{graphicx}
\usepackage{textcomp}
\usepackage{xcolor}
\usepackage[a4paper, total={184mm,239mm}]{geometry}

\usepackage{url}
\usepackage{booktabs}
\usepackage{tabularx}
\usepackage{array}
\usepackage{threeparttable}

\def\BibTeX{{\rm B\kern-.05em{\sc i\kern-.025em b}\kern-.08em
    T\kern-.1667em\lower.7ex\hbox{E}\kern-.125emX}}

\usepackage{tikz}
\usetikzlibrary{arrows.meta,calc,positioning,patterns}
\usetikzlibrary{arrows.meta,calc,positioning,patterns}
\definecolor{QBSBlue}{HTML}{24648C}
\definecolor{QBSTeal}{HTML}{147D70}
\definecolor{QBSGold}{HTML}{C08925}
\definecolor{QBSGray}{HTML}{626A73}
\definecolor{QBSRed}{HTML}{A44949}
\tikzset{paperbox/.style={draw=black!65,line width=.55pt,fill=white,
  align=center,inner sep=3pt,font=\fontsize{8.5}{10}\selectfont},
  route/.style={-{Latex[length=1.5mm,width=.95mm]},line width=.65pt},
  control/.style={route,dashed,draw=QBSGray},
  note/.style={font=\fontsize{8}{9.5}\selectfont,align=center},
  paneltitle/.style={anchor=west,font=\fontsize{8.5}{10}\selectfont\bfseries},
  junction/.style={circle,fill=black,inner sep=1.1pt}}

\begin{document}

% ============================================================
% Title and anonymous author information
% ============================================================

\title{Native Block-Quantized Matrix Multiplication and Explicit K/V Residency with Dual-View Access for LLM Inference on RISC-V Vector Processors}

\author{
	\IEEEauthorblockN{Weiying Wang, Yuanyang Xiang and Zhiwei Zhang}
	\IEEEauthorblockA{
		Institute of Automation, Chinese Academy of Sciences\\
		School of Artificial Intelligence, University of Chinese Academy of Sciences\\
		Beijing 100049, China\\
		Email: \{wangweiying2021,xiangyuanyang2020,zhiwei.zhang\}@ia.ac.cn
	}
}

\maketitle

% ============================================================
% Abstract
% ============================================================

\begin{abstract}
Some key execution states in large language model inference persist across multiple instructions and are difficult to express and exploit within RVV semantics centered on individual operations. We define specialization boundaries according to data-state lifetimes and propose Quantized Block Streams (QBS) and Attention/KV Context (AKV) while preserving RVV programmability. QBS organizes native-block format processing, block traversal, integer computation, and ordered FP32 accumulation into a unified execution process, whereas AKV maintains local K/V residency and supplies operand views aligned with the computation direction while leaving Attention arithmetic on RVV. Cycle-accurate RTL evaluation shows that QBS achieves geometric-mean speedups of 35.68$\times$ and 51.34$\times$ for projection \emph{Compute} and \emph{Matmul}, respectively, while AKV achieves a 1.45$\times$ geometric-mean speedup over Tiled RVV for Decode Attention. Compared with the baseline, the complete design increases total area by 12.65\%; on the evaluated projection workloads, throughput per area and energy efficiency improve by 46.63$\times$ and 19.78$\times$, respectively. Further evaluation demonstrates applicability across multiple block-quantization formats and across models with different Attention organizations.
\end{abstract}


% ============================================================
% I. Introduction
% ============================================================

\section{Introduction}
\label{sec:introduction}

Large language models (LLMs) are widely used in language understanding, content generation, and code processing~\cite{gpt3,llama2,codellama}, placing increasing demands on inference efficiency. Transformer inference repeatedly executes linear projections and attention operations~\cite{transformer}; as model sizes and context lengths continue to grow, the costs of weight access, attention computation, and intermediate-data movement become increasingly significant, making inference efficiency dependent on both computational throughput and data-supply efficiency. The RISC-V Vector Extension (RVV) provides vector-length-agnostic computation and memory operations, offering a unified programmable execution substrate across different hardware implementations and tensor shapes~\cite{rvv_spec}. However, runtime-known data layouts, reuse relationships, and state lifetimes are still largely managed by software through low-level vector operations, making it difficult for hardware to directly exploit data-organization and reuse patterns that persist across instructions. Based on this observation, we preserve RVV programmability while defining specialization boundaries according to LLM workload structure and optimizing the corresponding critical execution stages.

Inference runtimes such as \texttt{llama.cpp} employ block quantization to reduce model-weight storage and bandwidth requirements~\cite{llamacpp}, but compact representation does not directly translate into efficient vector execution. In addition to low-bit values, block-quantized weights contain scales, subgroup metadata, and format-specific correction rules, requiring RVV kernels to explicitly perform parsing, integer computation, correction, floating-point updates, and block traversal. Prior work optimizes quantized projection at different execution boundaries, including matrix-level execution, block-scaled dot products, and native quantized-block processing~\cite{mx,vmxdotp,spacemit_ime,fbfq,deca}; however, native block-quantized projection still requires format parsing, block traversal, and output accumulation to be coordinated with matrix-level reuse. Attention presents a different data-supply problem: K/V data are generated at runtime and repeatedly reused by subsequent queries, while grouped-query attention (GQA) and Prefill further enlarge their reuse scope~\cite{gqa}. Although methods such as FlashAttention reduce external data movement through tiling and online Softmax~\cite{flashattention}, local K/V residency organization and operand supply matched to the computation direction still require explicit software orchestration on vector processors.

\begin{figure*}[t]
\centering
\scalebox{0.80}{%
\begin{tikzpicture}[x=1cm,y=1cm,font=\fontsize{8.5}{10}\selectfont]
\path[use as bounding box] (0,.20) rectangle (18.3,3.10);

\node[paneltitle,text=black] at (0,2.93) {(a) Projection: output lifetime};
\node[paneltitle,text=black] at (9.45,2.93) {(b) Attention: operand lifetime};

\draw[black!20] (9.12,.08) -- (9.12,2.97);

\node[note,anchor=west] at (0,2.50) {Q4\_K: 256 weights};
\foreach \x/\w/\s in {3.70/1.65/{Scales 4 B},5.58/1.80/{Meta 12 B},7.64/2.14/{INT4 128 B}}{
  \node[paperbox,minimum width=\w cm,minimum height=.40cm,inner sep=2pt] at (\x,2.50) {\s};
}

\node[anchor=west,font=\bfseries\fontsize{8.5}{10}\selectfont] at (0,1.74) {RVV};
\foreach \x/\name/\label in {1.71/unpack/Unpack,3.48/mac/{Widen / dot},5.28/reduce/Reduce,7.34/scale/{Scale / update}}{
  \node[paperbox,minimum width=1.44cm,minimum height=.45cm,inner sep=2pt] (\name) at (\x,1.74) {\label};
}
\draw[route] (unpack) -- (mac);
\draw[route] (mac) -- (reduce);
\draw[route] (reduce) -- (scale);
\draw[control] (scale.south) -- (7.34,1.23) -- (1.71,1.23) -- (unpack.south);
\node[note,fill=white,inner sep=1pt] at (4.49,1.23) {Software K-block loop};

\node[anchor=west,font=\bfseries\fontsize{8.5}{10}\selectfont,text=QBSBlue] at (0,.65) {QBS};
\node[paperbox,draw=QBSBlue,fill=QBSBlue!5,minimum width=5.1cm,minimum height=.46cm] (native) at (3.54,.65)
  {Block stream $\rightarrow$ ordered accumulation};
\node[paperbox,draw=QBSBlue,minimum width=1.7cm,minimum height=.46cm] (out) at (7.69,.65) {FP32 tile};
\draw[route,QBSBlue] (native) -- (out);

\foreach \base/\name in {9.70/K,11.65/V}{
  \node[font=\bfseries\fontsize{8.5}{10}\selectfont] at (\base+.57,2.45) {\name};
  \foreach \r in {0,...,3}{\foreach \c in {0,...,4}{
    \fill[black!7] (\base+.23*\c,2.14-.19*\r) rectangle +(.22,.18);
  }}
}
\foreach \r in {0,...,3}{\fill[QBSTeal!70] (10.16,2.14-.19*\r) rectangle +(.22,.18);}
\foreach \c in {0,...,4}{\fill[QBSGold!75] (11.65+.23*\c,1.95) rectangle +(.22,.18);}

\node[note] at (11.25,1.32) {Token axis $t$ \quad Feature axis $d$};

\node[paperbox,draw=QBSTeal,text width=4.16cm,minimum width=4.40cm,minimum height=.74cm] (views) at (15.95,2.10)
 {K column $K[:,d]$ $\rightarrow$ QK\\V row $V[t,:]$ $\rightarrow$ PV};
\draw[route,QBSTeal] (12.95,1.93) -- (13.35,1.93) -- (13.35,2.1) -- (views.west);

\node[note] at (15.92,1.32) {RVV: QK $\rightarrow$ Softmax $\rightarrow$ PV};

\foreach \x/\n/\s/\w in {10.18/full/FILL/1.25,12.48/head/{Head / query reuse}/2.7,15.15/refill/REFILL/1.50,17.32/rel/RELEASE/1.55}{
  \node[paperbox,draw=QBSTeal,minimum width=\w cm,minimum height=.44cm,inner sep=2pt] (\n) at (\x,.65) {\s};
}
% FULL establishes the context, then consumers reuse the resident K/V.
\draw[route,QBSTeal] (full) -- (head);

% More K/V tiles: refill the context.
\draw[route,QBSTeal] (head) -- (refill);

% REFILL preserves the context and returns to the reuse phase.
\draw[control,QBSTeal]
  (refill.north) -- (15.15,1.08)
  -- (12.48,1.08) -- (head.north);

% Final use: release the resident context.
\draw[route,QBSTeal]
  (head.south) -- (12.48,.24)
  -- (17.32,.24) -- (rel.south);

\end{tikzpicture}
}
\caption{Specialization for cross-instruction lifetimes. QBS retains output state across $K$ blocks; AKV retains K/V across queries and supplies operand views.}
\label{fig:motivation}
\end{figure*}
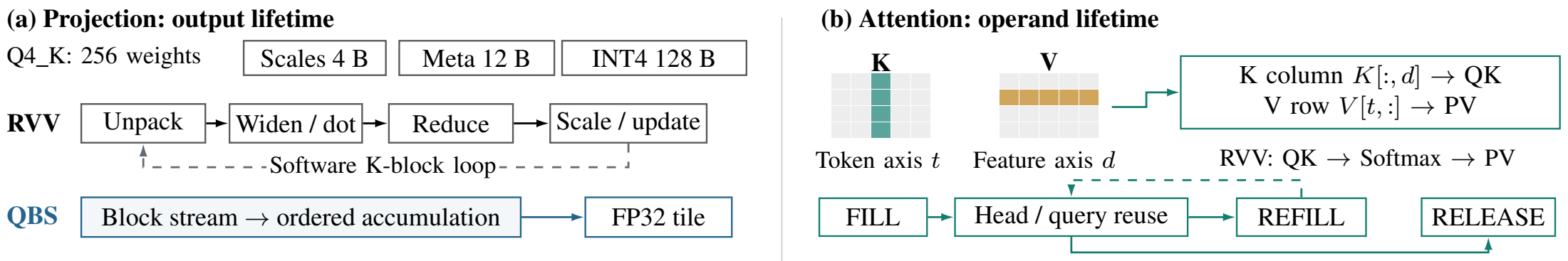

Although these two bottlenecks arise from different execution paths, they share a common property: the lifetimes of key data states extend beyond what a single RVV instruction can express. As shown in Fig.~\ref{fig:motivation}, quantized projection must preserve and update output accumulation state across successive K blocks, whereas attention must reuse K/V across queries and form different operand views according to the computation direction. Building on this observation, we propose \emph{Quantized Block Streams} (QBS) and an \emph{Attention/KV Context} (AKV). QBS defines a unified execution boundary spanning native quantized blocks to completed result tiles, bringing format handling, block traversal, and ordered accumulation across K blocks into the same execution process; AKV instead confines specialization to local K/V residency and operand supply, providing operand views matched to the computation direction. Attention arithmetic and other operations outside the supported scope continue to execute on standard RVV.

The main contributions of this work are as follows:
\begin{itemize}

    \item Specialization boundaries and architectural interfaces for cross-instruction state lifetimes.
    For ordered accumulation across native blocks in quantized projection and K/V reuse across attention queries, two architectural interfaces are defined from native quantized blocks to completed result tiles and from resident K/V to operand views, respectively.

    \item Integration and runtime mapping in RISC-V vector processors.
    QBS and AKV reuse the existing dependency-management, address-translation and memory-protection, and VRF result-delivery paths. The runtime selects specialized execution according to format, layout, shape, and data type, while operations outside the supported interface constraints continue to execute on standard RVV.

    \item RTL performance evaluation and functional validation with real-model workloads.
    QBS achieves geometric-mean speedups of 35.68$\times$ and 51.34$\times$ for projection Compute and Matmul, respectively, while AKV achieves a 1.45$\times$ geometric-mean speedup over GQA-aware Tiled RVV for Decode Attention. Validation covers multiple quantization formats and models with different Attention configurations.

\end{itemize}

% ============================================================
% II. Workload Structure and Related Work
% ============================================================

\section{Workload Structure and Related Work}
\label{sec:workload_related}

\subsection{Quantized Projection and K/V Reuse}
\label{sec:workload}

A linear projection computes $Y=AW^{\mathsf T}$, where $A\in\mathbb{R}^{M\times K}$ is the activation matrix and $W\in\mathbb{R}^{N\times K}$ is the weight matrix. Single-token Decode typically corresponds to $M=1$, whereas Prefill contains multiple activation rows. Block quantization partitions weights along the $K$ dimension into native blocks, so each output element $Y[m,n]$ is successively updated by multiple native blocks along $K$. As a concrete example, a Q4\_K native block occupies 144 bytes, represents 256 weights, and is divided into eight 32-element subgroups. When paired with Q8\_K activations, the integer computation within one block can be summarized by the dot-product and correction terms~\cite{llamacpp}

\begin{equation}
T=\sum_g s_g\sum_{i\in\mathcal{I}_g}q_i^a q_i^w,\qquad
U=\sum_g\mu_g\sum_{i\in\mathcal{I}_g}q_i^a .
\label{eq:q4k_terms}
\end{equation}

where $q_i^a$ and $q_i^w$ are the activation and weight integer values, respectively, while $s_g$ and $\mu_g$ encode subgroup scale and minimum information. Let $y$ denote the output accumulated from preceding native blocks. The current block then performs the format-defined FP32 update:

\begin{equation}
\begin{aligned}
z &= \operatorname{fma}_{32}\!\left(R_{32}(d_w d_a),R_{32}(T),y\right),\\
y' &= \operatorname{fma}_{32}\!\left(-R_{32}(d_{\min} d_a),R_{32}(U),z\right).
\end{aligned}
\label{eq:q4k_update}
\end{equation}

Here, $d_a$, $d_w$, and $d_{\min}$ denote the activation scale, weight scale, and minimum-value scale, respectively, and $R_{32}$ denotes the prescribed conversion or rounding to FP32. A native-block format therefore specifies not only the data encoding, but also the subgroup organization, scaling and correction rules, and block-level update order; beyond the low-precision dot product, its processing involves format parsing, format-specific correction, and FP32 updates. Matrix structure further exposes reuse in two directions: one activation block can serve multiple output channels along $N$, while one weight block can serve multiple activation rows along $M$. In the remainder of this paper, \emph{profile} denotes the encoding and numerical rules of a native block, \emph{layout} denotes its physical placement and traversal order, and \emph{shape} denotes the $M$, $N$, and $K$ range covered by one execution.

For Attention, let $Q\in\mathbb{R}^{M_q\times D}$ and $K,V\in\mathbb{R}^{T\times D}$. The computation is

\begin{equation}
P=\operatorname{softmax}\!\left(\frac{QK^{\mathsf T}}{\sqrt{D}}+\mathcal{M}\right),\qquad O=PV,
\label{eq:attention}
\end{equation}

where $D$ is the head dimension, $T$ is the number of currently visible K/V tokens, and $\mathcal{M}$ is the attention mask. In GQA, multiple Query heads share the same K/V head~\cite{gqa}, while Prefill allows the same K/V data to be reused by multiple Query positions. Fig.~\ref{fig:motivation}(b) illustrates this reuse and the corresponding operand-access directions. Under the vectorized schedule used in this work, attention-score computation fixes a feature dimension $d$ and reads $K[:,d]$ along the token dimension, whereas weighted accumulation over $V$ fixes a token $t$ and reads $V[t,:]$ along the feature dimension. Consequently, K columns and V rows are operand views formed by the vectorized schedule rather than inherent storage forms of K/V.

\subsection{Specialization Boundaries in Prior Work}
\label{sec:related_work}

Prior work on quantized projection and Attention explores specialization across computation granularity, data representation, execution mapping, and storage management~\cite{smoothquant,awq,kvquant}. For quantized projection, MX, VMXDOTP, and SpacemiT IME extend low-precision computation at the matrix, microscaling-dot-product, and tile levels, while F-BFQ and DECA combine native-format processing or near-core weight expansion with specialized matrix execution~\cite{mx,vmxdotp,spacemit_ime,fbfq,deca}. For Attention, FlashAttention and its successors primarily optimize tiling, online Softmax, and vector mapping~\cite{flashattention,flashattention2,flashattention3,vectorized_flashattention,flashattentionv}, whereas KIVI, PagedAttention, and PLENA reduce overhead through K/V representation, cache management, and on-chip data organization with dedicated execution support~\cite{kivi,pagedattention,plena}.

This work instead defines specialization boundaries from the common property of cross-instruction state lifetimes. QBS brings the path from native-block access to completed result-tile commit into a single specialized execution, jointly performing format parsing and correction, block traversal, and ordered accumulation across $K$ blocks; AKV confines specialization to local K/V residency and operand-view supply, while Attention arithmetic remains on standard RVV. The two mechanisms exploit output-accumulation state across $K$ blocks and K/V state across queries, respectively, leveraging cross-instruction state lifetimes and reuse relationships while preserving RVV programmability.

\section{Architectural Interfaces and Runtime Mapping}
\label{sec:interfaces}

QBS and AKV define their respective software-visible interfaces. Table~\ref{tab:instructions} summarizes the major operations, while Table~\ref{tab:limits} gives the range of a single execution.

\begin{table}[t]
\centering
\caption{Custom vector instructions of the proposed architecture.}
\label{tab:instructions}
\footnotesize
\setlength{\tabcolsep}{3pt}
\renewcommand{\arraystretch}{1.03}
\begin{tabularx}{0.96\columnwidth}{
@{}
>{\raggedright\arraybackslash}p{0.20\columnwidth}
>{\raggedright\arraybackslash}X
@{}}
\toprule
\textbf{Instruction} & \textbf{Function} \\
\midrule
\texttt{vqbinfo}     & Report supported profile, layout, and shape \\
\texttt{vqbexec}     & Execute native blocks to an $M\times N$ FP32 tile; support activation-snapshot reuse \\
\texttt{vakvinfo}    & Report supported context and view parameters \\
\texttt{vakvfill}    & Establish or update a K/V context (Fill/REFILL) \\
\texttt{vakvload.v}  & Load a selected Q/K/V row view to the VRF \\
\texttt{vakvkcol.v}  & Load a K-column or Panel4 view to the VRF \\
\texttt{vakvrelease} & Release the current K/V context \\
\bottomrule
\end{tabularx}
\end{table}

\begin{table}[t]
\centering
\caption{Architectural execution limits.}
\label{tab:limits}
\footnotesize
\setlength{\tabcolsep}{3pt}
\renewcommand{\arraystretch}{0.98}
\begin{tabular}{@{\hspace{2pt}}lccc@{\hspace{2pt}}}
\toprule
QBS Mode & M & N & K blocks \\
\midrule
Standard & 1--4 & $\leq32$ & $\leq256$ \\
Wide & 5--8 & $\leq16$ & $\leq256$ \\
Snapshot & 1 & $\leq32$ & $\leq16$ (Q8\_K) \\
\midrule
AKV Type & D & Q Rows & Tokens \\
\midrule
F16 K/V & 64/96/128 & 1--8 & $\leq64$ \\
\bottomrule
\end{tabular}
\end{table}

\subsection{QBS: Native Quantized-Block Execution Interface}
\label{sec:qbs_interface}

For $Y=AW^{\mathsf T}$, one QBS command processes $M$ activation rows and $N$ outputs, traverses a specified number of native quantized blocks along the $K$ dimension, and retains the output accumulation state until the reduction completes. The QBS descriptor specifies the weight address, profile, layout, $N$, and the number of K blocks, while the command additionally specifies the activation address, $M$, and the destination register. When $M$ or $N$ exceeds the command range, the runtime generates multiple output tiles along the corresponding dimension; when the number of K blocks exceeds the limit, split-K is used, with each K segment independently accumulated and rounded by a QBS command before software combines the partial results, so the combined result is not required to be bitwise identical to an unsplit execution.

When multiple output tiles share the same quantized activations, software can explicitly manage an activation snapshot through FILL, REUSE, and RELEASE. For Q8\_K activations satisfying the snapshot constraints, FILL saves the activation data read by the current command, REUSE makes the snapshot available to subsequent commands, and RELEASE frees it after the final successful use. The snapshot stores only already-quantized data, while F32-to-Q8\_K quantization remains a software responsibility; reuse is explicitly specified by software rather than inferred from the source address and changes only the cross-command data-supply path without altering the block-level computation or rounding rules of QBS.

\subsection{AKV: K/V Context and View Interface}
\label{sec:akv_interface}

Unlike QBS, which maintains output accumulation state within a command, AKV maintains K/V residency that can be reused across instructions through an explicit Attention context whose lifetime is managed using Fill, REFILL, and RELEASE. Fill establishes a new context and loads the first K/V tile, REFILL retains the existing Query data and context configuration while replacing only the currently resident K/V, and RELEASE frees the context after reuse completes. Once a context is established, software obtains the operands required by subsequent RVV computation through view operations: a row view returns the contiguous feature vector of a selected Query, K, or V row, a column view fixes one feature dimension and forms data arranged along the token direction, and Panel4 forms the column views corresponding to four adjacent feature dimensions in a single request. After these operands are written to the VRF, QK, Softmax, and PV continue to execute using standard RVV instructions; AKV therefore changes only K/V residency and operand supply rather than taking over Attention arithmetic. When the head dimension, K/V sequence length, or the number of Queries sharing the same K/V exceeds the range of a single execution, the runtime applies segmented scheduling along the feature dimension, partitions the K/V sequence into consecutive tiles along the token dimension, or groups Queries, respectively; Softmax state across K/V tiles and scheduling across segments and Query groups remain under software control.

\subsection{Runtime Composition and Fallback}
\label{sec:runtime_mapping}

QBS and AKV target quantized projection and K/V operand supply for Attention, respectively, and operate together with standard RVV to form the Transformer execution path. Q/K/V projections, output projections, and feed-forward matrix multiplications that satisfy the interface constraints can be mapped to QBS; after K/V cache writes complete and the AKV interface constraints are satisfied, AKV provides locally resident data and the required views for subsequent Attention computation, while the remaining operations continue to execute on RVV. Before dispatch, the runtime selects the specialized or RVV path according to constraints including quantization format, data layout, tensor shape, and data type, and directly uses the existing RVV implementation when the interface constraints are not satisfied.

\input{figures/architecture.tex}
\input{figures/datapaths.tex}

% ============================================================
% IV. Microarchitecture
% ============================================================

\section{Microarchitecture}
\label{sec:microarchitecture}

\subsection{Shared Execution, Dependencies, and State Visibility}
\label{sec:shared_execution}

Fig.~\ref{fig:architecture} presents the overall microarchitecture of QBS and AKV and their shared integration with the existing vector execution path. QBS, AKV, and ordinary vector memory operations share the VLSU address-translation path, PMA checks, AXI interface, and Lane result path, while the Sequencer continues to use existing instruction identifiers to track destination-register dependencies and completion state. QBS commands and AKV context operations wait for older vector operations to complete and for relevant ordinary memory transactions to drain before entering execution, and use the shared VLSU resources under mutually exclusive arbitration, ensuring that read responses, fault information, and returned results remain associated with the active execution request.

QBS result-accumulation state belongs only to the current command and is independent of the activation snapshot reused across commands; AKV K/V state may persist across multiple view instructions, while each view instruction retains its own destination register, dependencies, and completion state. Both mechanisms update architecturally visible state only after successful execution: QBS commits its final results after the associated memory accesses and computation complete, while AKV marks a context valid only after it has been completely filled without a fault. During reuse, QBS activation snapshots and AKV contexts do not reaccess the corresponding source addresses or automatically track changes to source memory, so software must rebuild or release the corresponding state when the source data change or reuse ends.

\subsection{QBS Native-Block Execution and Ordered Accumulation}
\label{sec:qbs_microarchitecture}

As shown in Fig.~\ref{fig:datapaths}(a), the QBS datapath performs native-block fetch, format decoding, integer computation, FP32 accumulation, and result commit. A tagged read path accommodates different native-block sizes and weight layouts; returned data are byte-aligned and reassembled into complete native quantized blocks before entering the decode path selected by the corresponding profile. The weight layout may be transformed during model loading; for example, the four-row block-interleaved layout (R4) interleaves native blocks at the same $K$ position from four output rows without changing their internal block encoding, thereby aligning the weight-read order with multi-output computation. Decoded data enter four parallel integer compute clusters, each containing eight signed INT8$\times$INT8 multipliers, for 32 integer multiplications in parallel. For $M=1,2,4$, each cluster serves one, two, or four activation rows, respectively; the Wide mode for $M=5$--$8$ reuses the same weight tile and processes the activation rows in two waves. Integer dot-product results are reduced within the corresponding subgroups and then corrected and scaled according to the active profile to form FP32 block-level updates. Different profiles primarily differ in format decoding and block-level update rules, while sharing the integer dot-product, reduction, and FP32 accumulation resources.

Block-level updates enter the FP32 accumulation path, where tags track in-flight updates and prevent conflicts on the same accumulation state. Updates for different outputs may proceed in an interleaved manner, while updates to the same output always preserve native $K$-block order. The current implementation provides 128 command-local FP32 accumulation states, allowing an entire $M\times N$ result tile to remain inside QBS throughout reduction along $K$ and avoiding repeated VRF writeback after each native block; after the reduction completes, the final FP32 results are written to the destination vector-register group.

\subsection{AKV Banked Storage and Multi-View Operand Supply}
\label{sec:akv_microarchitecture}

As shown in Fig.~\ref{fig:datapaths}(b), AKV distributes the resident K/V tile across eight single-port banks according to the token index, with token $t$ mapped to bank $t\bmod 8$ and K and V occupying separate logical regions while preserving the external row-major layout. This organization directly supports reading K/V rows by token and forming K column views by feature dimension. For a K-column access, the hardware fixes one feature dimension; in each round, the eight banks read the storage words corresponding to the current 8-token group and extract the F16 element at the same position, so a column containing $T_{\mathrm{valid}}$ valid tokens requires only $\lceil T_{\mathrm{valid}}/8\rceil$ bank-read rounds. Panel4 further extracts the F16 fields corresponding to four adjacent feature dimensions within the same read round and forms four column views without requiring additional SRAM read ports. K/V fill, row reads, and column-view generation share the same single-port banks under mutually exclusive arbitration; validity bits mask invalid tokens in tail tiles.

% ============================================================
% V. Evaluation
% ============================================================

\section{Evaluation}
\label{sec:evaluation}

\subsection{Evaluation Methodology and Baselines}
\label{sec:eval_method}

We use a functionally validated Ara implementation~\cite{ara2} with known implementation issues corrected as the baseline. All experiments use a four-lane configuration with VLEN=1024, ELEN=64, and a 128-bit AXI interface. Operator-level tests use cycle-accurate RTL simulation with tensors captured from real-model execution. Paired experiments use identical inputs and numerical acceptance criteria, with data layouts organized according to the corresponding kernels. We report execution cycles, AXI response bytes on the shared vector port, and retired vector instruction counts, and use internal hardware counters to analyze the major execution constraints.

For projection, the baseline uses format-specific multi-output RVV kernels optimized on Ara. \emph{Compute} includes dynamic activation quantization and matrix computation, whereas \emph{Matmul} denotes the matrix-computation phase; static weight prepacking is performed during model loading. For Attention, we compare Direct RVV, GQA-aware Tiled RVV, and AKV. Tiled RVV already applies software tiling and K/V reuse across queries, so its comparison with AKV measures the incremental benefit of local K/V residency and operand-view supply. Model-level functional runs validate runtime dispatch and fallback across different models and Attention configurations. PPA evaluation uses the same architectural parameters and implementation constraints to quantify the implementation cost and efficiency of the proposed design.

% Fig. 4
\input{figures/results.tex}

% ============================================================
% 5.2 Quantized Projection Performance and Reuse
% ============================================================

\subsection{Quantized Projection Performance and Reuse}
\label{sec:eval_projection}

Fig.~\ref{fig:results}(a) compares projection performance across representative block-quantized formats. QBS achieves geometric-mean speedups of 35.68$\times$ and 51.34$\times$ for Compute and Matmul, respectively, showing that organizing format processing, integer computation, and result updates as native-block execution substantially reduces quantized-projection overhead. Fig.~\ref{fig:results}(c) further supports this result from both data movement and instruction execution: all evaluated formats return fewer AXI bytes than the RVV baseline, while the number of retired vector instructions falls to only 0.42\%--0.56\% of the baseline. The performance gain therefore reflects not only a more efficient compute path, but also reduced data movement and explicit RVV instruction overhead.

\begin{figure}[t]
\centering
\scalebox{0.75}{%
\begin{tikzpicture}[
  x=1cm,
  y=1cm,
  font=\fontsize{8}{9.5}\selectfont,
  rnote/.style={note,inner sep=1pt},
  rtitle/.style={paneltitle,inner sep=0pt},
  tile/.style={draw=QBSBlue,fill=QBSBlue!5,line width=.55pt},
  weight/.style={draw=QBSBlue,fill=QBSBlue!15,line width=.55pt}
]

\path[use as bounding box] (0,0) rectangle (8.90,4.30);

% =========================================================
% Panel titles
% =========================================================

\node[rtitle,text=QBSBlue] at (.06,4.15)
  {(a) QBS payload bytes};

\node[rtitle,text=QBSTeal] at (4.72,4.15)
  {(b) AKV bank-access rounds};
  
% Divider moved slightly right to give QBS more width.
\draw[black!20]
  (4.60,.04) -- (4.60,4.27);

% =========================================================
% (a) QBS: reuse mechanism
% =========================================================

\node[
  rnote,
  text=QBSGray,
  font=\fontsize{8}{9.5}\selectfont\bfseries
] at (1.12,3.62)
  {Two M4};

\node[
  rnote,
  text=QBSBlue,
  font=\fontsize{8}{9.5}\selectfont\bfseries
] at (3.35,3.62)
  {One M8};

% ---------------------------------------------------------
% Two M4
% ---------------------------------------------------------

\foreach \y/\rows in {
  3.15/{M0--3},
  2.68/{M4--7}
} {

  \draw[
    draw=QBSGray,
    fill=QBSGray!12,
    line width=.55pt
  ]
    (.13,\y-.17)
    rectangle
    (.79,\y+.17);

  \node[rnote,text=QBSGray]
    at (.46,\y)
    {Q4};

  \draw[
    draw=QBSGray,
    fill=QBSGray!5,
    line width=.55pt
  ]
    (1.07,\y-.19)
    rectangle
    (2.04,\y+.19);

  \node[rnote]
    at (1.555,\y)
    {\rows};

  \draw[route,QBSGray]
    (.79,\y)
    -- (1.07,\y);
}

% ---------------------------------------------------------
% One M8
% ---------------------------------------------------------

% Shifted slightly right to use the additional QBS width.
\draw[weight]
  (2.47,2.74)
  rectangle
  (3.13,3.08);

\node[rnote,text=QBSBlue]
  at (2.80,2.91)
  {Q4};

\draw[tile,fill=QBSBlue!10]
  (3.45,2.43)
  rectangle
  (4.42,3.39);

\path[
  pattern=north east lines,
  pattern color=QBSBlue!70
]
  (3.46,2.44)
  rectangle
  (3.57,3.38);

\draw[QBSBlue!60]
  (3.45,2.91)
  -- (4.42,2.91);

\node[rnote]
  at (3.98,3.15)
  {M0--3};

\node[rnote]
  at (3.98,2.68)
  {M4--7};

\draw[QBSBlue,line width=.65pt]
  (3.13,2.91)
  -- (3.28,2.91);

\draw[route,QBSBlue]
  (3.28,2.91)
  -- (3.28,3.15)
  -- (3.45,3.15);

\draw[route,QBSBlue]
  (3.28,2.91)
  -- (3.28,2.68)
  -- (3.45,2.68);

\node[rnote,text=QBSGray]
  at (1.12,2.24)
  {2 reads};

\node[rnote,text=QBSBlue]
  at (3.35,2.24)
  {1 read};

% =========================================================
% (a) QBS: payload result
% =========================================================
%
% Four-column layout:
%
% Payload     Two M4     One M8        Red.
%              2.22       3.28         4.40
%
% The payload-label column is intentionally kept compact.
% =========================================================

\node[
  rnote,
  anchor=west,
  text=QBSGray
] at (.10,1.68)
  {Payload};

\node[
  rnote,
  anchor=east,
  text=QBSGray
] at (2.22,1.68)
  {Two M4};

\node[
  rnote,
  anchor=east,
  text=QBSBlue
] at (3.34,1.68)
  {One M8};

\node[
  rnote,
  anchor=east,
  text=QBSBlue
] at (4.25,1.68)
  {Red.};

\draw[black!30]
  (.18,1.50)
  -- (4.40,1.50);

% ---------------------------------------------------------
% Q8_K payload
% ---------------------------------------------------------

\node[
  rnote,
  anchor=west
] at (.18,1.26)
  {Q8\_K};

\node[
  rnote,
  anchor=east
] at (2.22,1.26)
  {2,336};

\node[
  rnote,
  anchor=east
] at (3.28,1.26)
  {2,336};

% ---------------------------------------------------------
% Q4_K payload
% ---------------------------------------------------------

\node[
  rnote,
  anchor=west
] at (.18,.95)
  {Q4\_K};

\node[
  rnote,
  anchor=east,
  text=QBSGray
] at (2.22,.95)
  {4,608};

\node[
  rnote,
  anchor=east,
  text=QBSBlue
] at (3.28,.95)
  {2,304};

\draw[black!30]
  (.18,.77)
  -- (4.40,.77);

% ---------------------------------------------------------
% Total + reduction
% ---------------------------------------------------------

\node[
  rnote,
  anchor=west,
  font=\fontsize{8.5}{10}\selectfont\bfseries
] at (.18,.54)
  {Total};

\node[
  rnote,
  anchor=east,
  text=QBSGray,
  font=\fontsize{8.5}{10}\selectfont\bfseries
] at (2.22,.54)
  {6,944};

\node[
  rnote,
  anchor=east,
  text=QBSBlue,
  font=\fontsize{8.5}{10}\selectfont\bfseries
] at (3.28,.54)
  {4,640};

\node[
  rnote,
  anchor=east,
  text=QBSBlue,
  font=\fontsize{8.5}{10}\selectfont\bfseries
] at (4.40,.54)
  {$-33.2\%$};

% =========================================================
% (b) AKV: access mechanism
% =========================================================

\node[rnote]
  at (6.78,3.63)
  {8 single-port banks; 256-bit words};

% ---------------------------------------------------------
% One bank word
% ---------------------------------------------------------

\draw[
  draw=QBSTeal,
  line width=.55pt
]
  (4.91,3.05)
  rectangle
  (8.65,3.39);

\node[rnote]
  at (5.14,3.22)
  {$\cdots$};

\node[rnote]
  at (8.42,3.22)
  {$\cdots$};

\foreach \i/\field in {
  0/d,
  1/{d{+}1},
  2/{d{+}2},
  3/{d{+}3}
} {

  \pgfmathsetmacro{\x}{5.37+.705*\i}

  \draw[
    draw=QBSTeal,
    fill=QBSTeal!16,
    line width=.55pt
  ]
    (\x,3.05)
    rectangle
    (\x+.705,3.39);

  \node[rnote]
    at (\x+.3525,3.22)
    {$\field$};
}

% ---------------------------------------------------------
% Four independent single-column accesses
% ---------------------------------------------------------

\draw[route,QBSGray]
  (5.49,3.05)
  -- (5.49,2.88)
  -- (4.85,2.88)
  -- (4.85,2.46)
  -- (5.14,2.46);

\node[
  rnote,
  text=QBSGray,
  font=\fontsize{8}{9.5}\selectfont\bfseries
] at (5.69,2.46)
  {4 singles};

\node[
  rnote,
  text=QBSGray
] at (5.69,2.11)
  {1 field/word};

% ---------------------------------------------------------
% Panel4 access
% ---------------------------------------------------------

\draw[route,QBSTeal]
  (8.07,3.05)
  -- (8.07,2.88)
  -- (8.72,2.88)
  -- (8.72,2.46)
  -- (8.43,2.46);

\node[
  rnote,
  text=QBSTeal,
  font=\fontsize{8}{9.5}\selectfont\bfseries
] at (7.88,2.46)
  {Panel4};

\node[
  rnote,
  text=QBSTeal
] at (7.88,2.11)
  {4 fields/word};

% =========================================================
% (b) AKV: bank-round result
% =========================================================

\foreach \i in {0,...,3} {

  \pgfmathsetmacro{\x}{4.82+.47*\i}

  \draw[
    draw=QBSGray,
    fill=QBSGray!12,
    line width=.55pt
  ]
    (\x,1.25)
    rectangle
    (\x+.38,1.61);

  \node[rnote]
    at (\x+.19,1.43)
    {8};
}

\draw[
  draw=QBSTeal,
  fill=QBSTeal!16,
  line width=.55pt
]
  (7.66,1.25)
  rectangle
  (8.10,1.61);

\node[rnote]
  at (7.88,1.43)
  {8};

\node[
  rnote,
  text=QBSGray,
  font=\fontsize{8}{9.5}\selectfont\bfseries
] at (5.69,.98)
  {32 rounds};

\node[
  rnote,
  text=QBSTeal,
  font=\fontsize{8}{9.5}\selectfont\bfseries
] at (7.88,.98)
  {8 rounds};

\draw[black!30]
  (4.73,.71)
  -- (8.77,.71);

\node[
  rnote,
  text=QBSTeal,
  font=\fontsize{8.5}{10}\selectfont\bfseries
] at (6.78,.49)
  {$-75\%$ bank rounds};

\end{tikzpicture}
}
\caption{Mechanism-level reuse in QBS and AKV: (a) Wide-M weight-payload reuse and (b) Panel4 bank-access reduction. Values represent data-supply work rather than execution latency.}
\label{fig:reuse}
\end{figure}
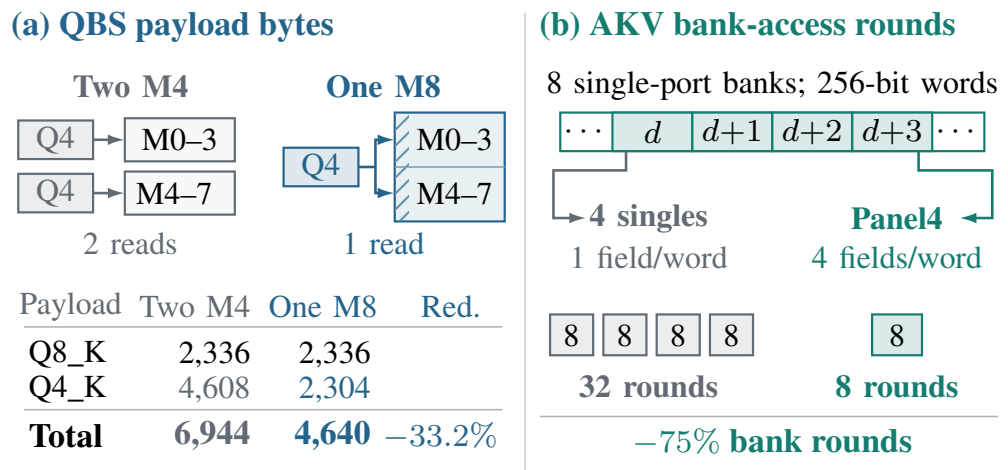

Fig.~\ref{fig:attribution}(a) further exposes format-dependent execution behavior. Q3\_K--Q6\_K are primarily dominated by dot-product execution, with more pronounced weight-supply stalls in Q5\_K and Q6\_K. In contrast, the more frequent result updates of Q8\_0 make FP32 update overhead more prominent. Fig.~\ref{fig:attribution}(b) and Fig.~\ref{fig:reuse}(a) examine activation and weight reuse, respectively. Reusing quantized activations reduces the total cycles of three identical projections by 47.6\%, while Wide-M shares weights across activation rows and reduces the input payload by 33.2\% for equal work. The former avoids repeated activation quantization, whereas the latter reduces redundant weight reads, further reducing projection overhead from activation preparation and weight supply.

% Fig. 5
\input{figures/attribution.tex}

% Fig. 6

% ============================================================
% 5.3 Attention Performance and K/V Supply
% ============================================================

\subsection{Attention Performance and K/V Supply}
\label{sec:eval_attention}

Fig.~\ref{fig:results}(b) compares Attention latency during Decode. Relative to Tiled RVV, which already exploits GQA-aware tiling and K/V reuse across queries, AKV achieves a 1.45$\times$ geometric-mean speedup. As the KV length increases from 16 to 256, the speedup rises from 1.22$\times$ to 1.60$\times$. Since the software baseline already explicitly exploits K/V sharing, the remaining gain shows that K/V organization and delivery still constitute significant data-supply overhead, whose impact becomes more pronounced as the context grows.

Fig.~\ref{fig:attribution}(c) further decomposes this gain. With the effective K/V payload unchanged, two-query sharing combined with Panel4 reduces column-view commands and local replay by 81.7\% and 30.4\%, respectively, while reducing execution cycles by 14.5\%. The improvement therefore primarily comes from more efficient organization and supply of resident K/V rather than from reducing the effective data volume or the Attention computation itself. Fig.~\ref{fig:reuse}(b) explains this behavior at the storage-access level: Panel4 exploits adjacent F16 fields already available in a single 256-bit bank word, reducing the bank-access rounds required for four adjacent column views from 32 to 8 while preserving the single-port bank organization. AKV thus confines specialization to K/V residency and operand-view supply, while QK, Softmax, and PV continue to execute on the existing RVV datapath.

\subsection{Model Applicability and Runtime Validation}
\label{sec:eval_generality}

This section evaluates design applicability through cross-configuration RTL tests and model-level functional runs. Beyond the D128/GQA configurations used in the main evaluation, Refact D64/MQA32~\cite{refact} and Gemma~3 D256/GQA4~\cite{gemma3} reuse the existing K/V residency and view-supply structures and achieve 2.04$\times$ and 2.55$\times$ speedups over Direct RVV, respectively. Model-level functional runs cover Qwen2.5~\cite{qwen25}, Qwen3~\cite{qwen3}, Refact, and Gemma~3; all cases execute native QBS commands and validate AKV runtime dispatch and fallback, and all pass functional checks. These results show that different model configurations can be accommodated through software scheduling without changing the core hardware structures, thereby supporting multiple model configurations while preserving compatibility with the existing execution path.

\subsection{Implementation Cost and Efficiency}
\label{sec:eval_ppa}

Table~\ref{tab:fixed_resource_eff} compares the PPA and efficiency results of baseline Ara and the proposed design under the same technology, processor configuration, and implementation constraints. At an operating frequency of 1\,GHz, the total area increases from 3.195\,mm$^2$ to 3.599\,mm$^2$, corresponding to a 12.65\% area overhead, while the overall performance reaches 51.34$\times$ that of the baseline and throughput per area improves by 46.63$\times$. Relative to the performance gain, the area overhead introduced by the additional compute, state-storage, and data-supply logic remains limited, indicating that the additional hardware resources can be effectively translated into processing throughput. The average power of the proposed design is 2.67$\times$ that of the baseline, while energy efficiency improves from 3.55 to 70.18\,GOPS/W, corresponding to a 19.78$\times$ gain. The performance improvement is substantially larger than the power increase, indicating that reducing the overhead of format processing, explicit vector execution, and operand supply can effectively amortize the additional hardware power cost and thereby deliver significant energy-efficiency gains.

% PPA and efficiency results
\begin{table}[t]
\centering
\caption{PPA and efficiency comparison.}
\label{tab:fixed_resource_eff}
\scriptsize
\setlength{\tabcolsep}{0pt}
\renewcommand{\arraystretch}{1.08}

\begin{threeparttable}
\begin{tabular*}{0.98\columnwidth}{@{\extracolsep{\fill}}lccc@{}}
\toprule
\textbf{Metric} &
\textbf{Baseline Ara} &
\textbf{This Work} &
\textbf{Relative Change} \\
\midrule

Technology
& \multicolumn{2}{c}{TSMC 28\,nm HPC+}
& -- \\

Frequency (GHz)
& \multicolumn{2}{c}{1.0}
& -- \\

Lanes / VLEN / DLEN
& \multicolumn{2}{c}{4 / 1024 / 256}
& -- \\

Performance (GOPS)
& 0.615
& 32.29
& 51.34$\times$ \\

Area (mm$^2$)
& 3.195
& 3.599
& +12.65\% \\

Area Eff. (GOPS/mm$^2$)
& 0.192
& 8.97
& 46.63$\times$ \\

Power (mW)
& 177.0
& 471.7
& +166.49\% \\

Execution Energy (pJ/op)
& 312.8
& 16.91
& $-$94.59\% \\

Energy Eff. (GOPS/W)
& 3.55
& 70.18
& 19.78$\times$ \\

EDP (norm.)
& 1.00
& 0.0010
& $-$99.90\% \\

\bottomrule
\end{tabular*}

\begin{tablenotes}[flushleft]
\scriptsize
\item[] \textit{Note:} Both designs use TSMC 28\,nm HPC+ (TT/0.9\,V/25$^\circ$C) at 1\,GHz with four lanes, VLEN=1024, and DLEN=256.
\end{tablenotes}

\end{threeparttable}
\end{table}

\section{Conclusion}
\label{sec:conclusion}

This paper proposes QBS and AKV to exploit state and reuse relationships that persist across instructions in LLM inference. QBS unifies native-block format processing, block traversal, and ordered FP32 accumulation, while AKV reduces operand-organization overhead through local K/V residency and multi-view supply while preserving the existing RVV arithmetic path. RTL evaluation shows that the two mechanisms reduce quantized-projection and K/V-supply overhead while improving throughput per area and energy efficiency. Validation across multiple quantization formats and model-specific Attention organizations further shows that the design accommodates diverse LLM inference configurations while retaining a unified hardware structure and RVV programmability.

% ============================================================
% References
% ============================================================

\clearpage
\bibliographystyle{IEEEtran}
\bibliography{references}

@manual{rvv_spec,
  author       = {{RISC-V International}},
  title        = {{RISC-V ``V'' Vector Extension, Version 1.0}},
  year         = {2021},
  organization = {RISC-V International},
  url          = {https://docs.riscv.org/reference/isa/extensions/vector/_attachments/riscv-v-spec.pdf}
}

@inproceedings{transformer,
  author    = {Ashish Vaswani and Noam Shazeer and Niki Parmar and Jakob Uszkoreit and
               Llion Jones and Aidan N. Gomez and Lukasz Kaiser and Illia Polosukhin},
  title     = {Attention Is All You Need},
  booktitle = {Advances in Neural Information Processing Systems},
  volume    = {30},
  year      = {2017}
}

@inproceedings{gpt3,
  author    = {Tom B. Brown and Benjamin Mann and Nick Ryder and Melanie Subbiah and
               Jared D. Kaplan and Prafulla Dhariwal and Arvind Neelakantan and
               Pranav Shyam and Girish Sastry and Amanda Askell and others},
  title     = {Language Models are Few-Shot Learners},
  booktitle = {Advances in Neural Information Processing Systems},
  volume    = {33},
  pages     = {1877--1901},
  year      = {2020}
}

@misc{llama2,
  author        = {Hugo Touvron and Louis Martin and Kevin Stone and Peter Albert and
                   Amjad Almahairi and Yasmine Babaei and Nikolay Bashlykov and
                   Soumya Batra and Prajjwal Bhargava and Shruti Bhosale and others},
  title         = {{Llama 2}: Open Foundation and Fine-Tuned Chat Models},
  year          = {2023},
  eprint        = {2307.09288},
  archivePrefix = {arXiv},
  primaryClass  = {cs.CL},
  url           = {https://arxiv.org/abs/2307.09288}
}

@misc{codellama,
  author        = {Baptiste Rozi{\`e}re and Jonas Gehring and Fabian Gloeckle and
                   Sten Sootla and Itai Gat and Xiaoqing Ellen Tan and Yossi Adi and
                   Jingyu Liu and Romain Sauvestre and Tal Remez and others},
  title         = {{Code Llama}: Open Foundation Models for Code},
  year          = {2023},
  eprint        = {2308.12950},
  archivePrefix = {arXiv},
  primaryClass  = {cs.CL},
  url           = {https://arxiv.org/abs/2308.12950}
}

@misc{llamacpp,
  author       = {Georgi Gerganov and contributors},
  title        = {{llama.cpp: LLM Inference in C/C++}},
  year         = {2026},
  howpublished = {\url{https://github.com/ggml-org/llama.cpp}},
  note         = {Accessed: 2026-09-19}
}

@inproceedings{gqa,
  author    = {Joshua Ainslie and James Lee-Thorp and Michiel de Jong and
               Yury Zemlyanskiy and Federico Lebron and Sumit Sanghai},
  title     = {{GQA}: Training Generalized Multi-Query Transformer Models
               from Multi-Head Checkpoints},
  booktitle = {Proceedings of the 2023 Conference on Empirical Methods
               in Natural Language Processing (EMNLP)},
  pages     = {4895--4901},
  year      = {2023},
  publisher = {Association for Computational Linguistics},
  doi       = {10.18653/v1/2023.emnlp-main.298}
}

@inproceedings{flashattention,
  author    = {Tri Dao and Daniel Y. Fu and Stefano Ermon and
               Atri Rudra and Christopher R{\'e}},
  title     = {{FlashAttention}: Fast and Memory-Efficient Exact Attention
               with {IO}-Awareness},
  booktitle = {Advances in Neural Information Processing Systems},
  volume    = {35},
  pages     = {16344--16359},
  year      = {2022}
}

@inproceedings{flashattention2,
  author    = {Tri Dao},
  title     = {{FlashAttention-2}: Faster Attention with Better Parallelism
               and Work Partitioning},
  booktitle = {The Twelfth International Conference on Learning Representations (ICLR)},
  year      = {2024},
  url       = {https://openreview.net/forum?id=mZn2Xyh9Ec}
}

@inproceedings{flashattention3,
  author    = {Jay Shah and Ganesh Bikshandi and Ying Zhang and Vijay Thakkar and
               Pradeep Ramani and Tri Dao},
  title     = {{FlashAttention-3}: Fast and Accurate Attention with Asynchrony
               and Low-Precision},
  booktitle = {Advances in Neural Information Processing Systems},
  volume    = {37},
  year      = {2024},
  doi       = {10.52202/079017-2193}
}

@inproceedings{smoothquant,
  author    = {Guangxuan Xiao and Ji Lin and Mickael Seznec and Hao Wu and
               Julien Demouth and Song Han},
  title     = {{SmoothQuant}: Accurate and Efficient Post-Training Quantization
               for Large Language Models},
  booktitle = {Proceedings of the 40th International Conference on Machine Learning},
  series    = {Proceedings of Machine Learning Research},
  volume    = {202},
  pages     = {38087--38099},
  publisher = {PMLR},
  year      = {2023},
  url       = {https://proceedings.mlr.press/v202/xiao23c.html}
}

@inproceedings{awq,
  author    = {Ji Lin and Jiaming Tang and Haotian Tang and Shang Yang and
               Wei-Ming Chen and Wei-Chen Wang and Guangxuan Xiao and Xingyu Dang and
               Chuang Gan and Song Han},
  title     = {{AWQ}: Activation-Aware Weight Quantization for On-Device
               {LLM} Compression and Acceleration},
  booktitle = {Proceedings of Machine Learning and Systems},
  volume    = {6},
  year      = {2024},
  url       = {https://proceedings.mlsys.org/paper_files/paper/2024/hash/42a452cbafa9dd64e9ba4aa95cc1ef21-Abstract-Conference.html}
}

@inproceedings{kvquant,
  author    = {Coleman Hooper and Sehoon Kim and Hiva Mohammadzadeh and
               Michael W. Mahoney and Yakun Sophia Shao and Kurt Keutzer and
               Amir Gholami},
  title     = {{KVQuant}: Towards 10 Million Context Length {LLM} Inference
               with {KV} Cache Quantization},
  booktitle = {Advances in Neural Information Processing Systems},
  volume    = {37},
  year      = {2024},
  doi       = {10.52202/079017-0040}
}

@inproceedings{mx,
  author    = {Matteo Perotti and Yichao Zhang and Matheus A. Cavalcante and
               Enis Mustafa and Luca Benini},
  title     = {{MX}: Enhancing {RISC-V}'s Vector {ISA} for Ultra-Low
               Overhead, Energy-Efficient Matrix Multiplication},
  booktitle = {2024 Design, Automation \& Test in Europe Conference
               \& Exhibition (DATE)},
  pages     = {1--6},
  year      = {2024},
  doi       = {10.23919/DATE58400.2024.10546720}
}

@inproceedings{vmxdotp,
  author    = {Max Wipfli and Gamze Islamoglu and Navaneeth Kunhi Purayil and
               Angelo Garofalo and Luca Benini},
  title     = {{VMXDOTP}: A {RISC-V} Vector {ISA} Extension for Efficient
               Microscaling ({MX}) Format Acceleration},
  booktitle = {2026 Design, Automation \& Test in Europe Conference (DATE)},
  pages     = {1--7},
  year      = {2026},
  doi       = {10.23919/DATE69613.2026.11539385}
}

@misc{spacemit_ime,
  author       = {{SpacemiT}},
  title        = {{RISC-V IME Set Specification}},
  year         = {2026},
  howpublished = {\url{https://github.com/spacemit-com/riscv-ime-extension-spec}},
  note         = {Version 1.0 specification; accessed: 2026-09-19}
}

@inproceedings{fbfq,
  author    = {Jude Haris and Jos{\'e} Cano},
  title     = {{F-BFQ}: Flexible Block Floating-Point Quantization
               Accelerator for {LLMs}},
  booktitle = {New Approaches for Addressing the Computing Requirements
               of LLMs and GNNs (LG-ARC), co-located with ISCA},
  address   = {Tokyo, Japan},
  year      = {2025},
  month     = jun,
  url       = {https://arxiv.org/abs/2510.13401}
}

@inproceedings{deca,
  author    = {Gerasimos Gerogiannis and Stijn Eyerman and
               Evangelos Georganas and Wim Heirman and Josep Torrellas},
  title     = {{DECA}: A Near-Core {LLM} Decompression Accelerator
               Grounded on a 3D Roofline Model},
  booktitle = {58th IEEE/ACM International Symposium on
               Microarchitecture (MICRO)},
  pages     = {184--200},
  year      = {2025},
  doi       = {10.1145/3725843.3756073}
}

@article{vectorized_flashattention,
  author  = {Vasileios Titopoulos and Kosmas Alexandridis and
             Giorgos Dimitrakopoulos},
  title   = {Vectorized {FlashAttention} with Low-Cost Exponential
             Computation in {RISC-V} Vector Processors},
  journal = {The Journal of Supercomputing},
  volume  = {82},
  number  = {4},
  pages   = {189},
  year    = {2026},
  doi     = {10.1007/s11227-026-08322-x}
}

@misc{flashattentionv,
  author        = {Sonia Rani Gupta and Nikela Papadopoulou and
                   Miquel Peric{\`a}s},
  title         = {{FlashAttention} for Scalable Vector Architectures},
  year          = {2026},
  eprint        = {2608.18656},
  archivePrefix = {arXiv},
  primaryClass  = {cs.AR},
  url           = {https://arxiv.org/abs/2608.18656}
}

@inproceedings{kivi,
  author    = {Zirui Liu and Jiayi Yuan and Hongye Jin and
               Shaochen Zhong and Zhaozhuo Xu and Vladimir Braverman and
               Beidi Chen and Xia Hu},
  title     = {{KIVI}: A Tuning-Free Asymmetric 2bit Quantization
               for {KV} Cache},
  booktitle = {Proceedings of the 41st International Conference on
               Machine Learning},
  series    = {Proceedings of Machine Learning Research},
  volume    = {235},
  pages     = {32332--32344},
  publisher = {PMLR},
  year      = {2024},
  url       = {https://proceedings.mlr.press/v235/liu24bz.html}
}

@inproceedings{pagedattention,
  author    = {Woosuk Kwon and Zhuohan Li and Siyuan Zhuang and
               Ying Sheng and Lianmin Zheng and Cody Hao Yu and
               Joseph E. Gonzalez and Hao Zhang and Ion Stoica},
  title     = {Efficient Memory Management for Large Language Model
               Serving with {PagedAttention}},
  booktitle = {Proceedings of the 29th Symposium on Operating Systems
               Principles (SOSP)},
  pages     = {611--626},
  publisher = {ACM},
  year      = {2023},
  doi       = {10.1145/3600006.3613165}
}

@inproceedings{plena,
  author    = {Haoran Wu and Can Xiao and Jiayi Nie and Xuan Guo and
               Binglei Lou and Jeffrey T. H. Wong and Zhiwen Mo and
               Cheng Zhang and Przemyslaw Forys and Chengyang Ai and
               Timi Adeniran and Wayne Luk and Hongxiang Fan and
               Jianyi Cheng and Timothy M. Jones and Rika Antonova and
               Robert Mullins and Aaron Zhao},
  title     = {Combating the Memory Walls: Optimization Pathways for
               Long-Context Agentic {LLM} Inference},
  booktitle = {Proceedings of the 53rd IEEE Annual International
               Symposium on Computer Architecture (ISCA)},
  pages     = {100--115},
  publisher = {IEEE},
  year      = {2026},
  doi       = {10.1109/ISCA66397.2026.00023}
}

@article{ara2,
  author  = {Matteo Perotti and Matheus Cavalcante and Renzo Andri and
             Lukas Cavigelli and Luca Benini},
  title   = {{Ara2}: Exploring Single- and Multi-Core Vector Processing
             With an Efficient {RVV} 1.0 Compliant Open-Source Processor},
  journal = {IEEE Transactions on Computers},
  volume  = {73},
  number  = {7},
  pages   = {1822--1836},
  year    = {2024},
  doi     = {10.1109/TC.2024.3388896}
}

@misc{qwen25,
  author        = {An Yang and others},
  title         = {{Qwen2.5} Technical Report},
  year          = {2024},
  eprint        = {2412.15115},
  archivePrefix = {arXiv},
  primaryClass  = {cs.CL},
  url           = {https://arxiv.org/abs/2412.15115}
}

@misc{qwen3,
  author        = {An Yang and others},
  title         = {{Qwen3} Technical Report},
  year          = {2025},
  eprint        = {2505.09388},
  archivePrefix = {arXiv},
  primaryClass  = {cs.CL},
  url           = {https://arxiv.org/abs/2505.09388}
}

@misc{gemma3,
  author        = {{Gemma Team}},
  title         = {{Gemma 3} Technical Report},
  year          = {2025},
  eprint        = {2503.19786},
  archivePrefix = {arXiv},
  url           = {https://arxiv.org/abs/2503.19786}
}

@misc{refact,
  author       = {{Refact.ai}},
  title        = {{Refact-1.6B-fim}},
  year         = {2023},
  howpublished = {Hugging Face model repository},
  url          = {https://huggingface.co/refactai/Refact-1_6B-fim},
  note         = {Accessed: 2026-09-21}
}

\end{document}